\documentclass[twocolumn,showpacs,preprintnumbers,amsmath,amssymb]{revtex4}

\usepackage{graphicx}
\usepackage{dcolumn}
\usepackage{bm}

\begin{document}

\preprint{}

\title{ Secure  direct communication using Einstein-Podolsky-Rosen pairs and teleportation}

\author{ Fengli Yan $^{1,2}$ and Xiaoqiang Zhang $^2$}

 \affiliation{  $^1$ CCAST (World Laboratory), P.O. Box 8730, Beijing 100080, China\\
$^2$ Department of Physics, Hebei Normal University, Shijiazhuang 050016, China\\}
\date{\today}
\begin{abstract}
A novel scheme for  secure direct communication between Alice and Bob is proposed, where there is no need for
establishing a shared secret key. The communication is based on Einstein-Podolsky-Rosen pairs and teleportation
between Alice and Bob. After insuring the security of the quantum channel (EPR pairs), Bob encodes the secret
message directly on a sequence of particle states  and transmits them  to Alice by teleportation. In this scheme
teleportation transmits  Bob's message without revealing any information to a potential eavesdropper. Alice can
read out the encoded messages directly by the measurement on her qubits. Because there is not a transmission of
the qubit which carry the secret message between Alice and Bob, it is completely secure for direct secret
communication if perfect quantum channel is used.
\end{abstract}

\maketitle

Since languages become the tool for communication, the desire and need to transmit secret messages from one
person to another begin. Then human have the cryptography - an art to transmit information so that it is
unintelligible and therefore useless to those who are not meant to have access to it. It is generally believed
that cryptography schemes are only completely secure when the two communicating parties, Alice and Bob,
establish a shared secret key before the transmission of a message. This means they first have to create a
random bit sequence, which is not known to anyone else, and which is of the same length as the message. In order
to communicate, Alice then multiplies the bits of the message one by one with the key bits. When she announces
the result to Bob, or even publicly, then he is the only one who can interpret it and deduce Alice's message.

But it is difficult to distribute securely the secret key through a classical channel. Quantum key distribution
(QKD), the approach using quantum mechanics principle for distribution of secret key is the only proven protocol
for secure key distribution.

Since Bennett and Brassard proposed the standard BB84 QKD protocol [1] in 1984, it has been developed quickly.
Up to now there have already been a lot of theoretical QKD schemes, for instance in Refs. \cite {s1, s2, s3, s4,
s5,  s6,  s7,  s8,  s9,  s10,  s11, s12, s13, s14, s15, s16, s17, s18}.

Different from key distribution whose object is to establish a common random key between two parties, a secure
direct communication is to communicate important messages directly without first establishing a random key to
encrypt them.  As classical message can be copied fully, it is impossible to transmit secret messages directly
through classical channels.  Recently Beige et al. \cite {s19} proposed a quantum secure direct communication
(QSDC) scheme. In this scheme the message can be read out only after a transmission of an additional classical
information for each qubit. Bostr\"{o}m and Felbinger put forward a Ping-Pong QSDC scheme [20]. It is secure for
key distribution, quasi-secure for direct secret communication if perfect quantum channel is used. But it is
insecure if it is operated in a noisy quantum channel, as shown by W\'{o}jcik [21]. There is some probability
that a part of the messages might be leaked to the eavesdropper, Eve, especially in a noisy quantum channel,
because Eve can use the intercept-resending strategy to steal some secret messages even though Alice and Bob
will find out her in the end of communication, especially in a noise quantum channel. More recently Deng et al
\cite {s22} proposed a two-step quantum direct communication protocol using Einstein-Podolsky-Rosen pair block.
It was shown that it is provably secure. However in all these secure direct communication schemes  it is
necessary to send  the
  qubits carrying  secret messages in the public channel. Therefore,  Eve can  attack the qubits  in
   transmission.

  In this paper we present a scheme for direct and confidential communication between Alice and Bob, where there is no need
   for establishing a shared secret key. The secure direct communication is based on Einstein-Podolsky-Rosen pairs and teleportation
    \cite {s23}.   Because  there is not a transmission of  the qubit which carry the secret message between Alice
and Bob in the public channel, it is completely secure for direct secret communication if perfect quantum
channel is used.

  The new protocol can be divided into two steps, one is to prepare   EPR pairs (quantum channel), the other is to
  transmit messages using  teleportation.

{\it Preparing  EPR pairs.} --- Suppose that Alice and Bob share a set of  entangled pairs of qubits
 in one of the Bell's states
  \begin{eqnarray*}
&& |\Phi^+\rangle_{AB}=\frac {1}{\sqrt 2}(|00\rangle_{AB}+|11\rangle_{AB})\\
&&~~~~~~~~~~ =\frac {1}{\sqrt
2}(|+\rangle_A|+\rangle_B+|-\rangle_A|-\rangle_B),\\
&& |\Phi^-\rangle_{AB}=\frac {1}{\sqrt 2}(|00\rangle_{AB}-|11\rangle_{AB})\\
&&~~~~~~~~~~ =\frac {1}{\sqrt
2}(|+\rangle_A|-\rangle_B+|-\rangle_A|+\rangle_B),\\
&& |\Psi^+\rangle_{AB}=\frac {1}{\sqrt 2}(|01\rangle_{AB}+|10\rangle_{AB})\\
&&~~~~~~~~~~ =\frac {1}{\sqrt
2}(|+\rangle_A|+\rangle_B-|-\rangle_A|-\rangle_B),\\
&& |\Psi^-\rangle_{AB}=\frac {1}{\sqrt 2}(|01\rangle_{AB}-|10\rangle_{AB})\\
&&~~~~~~~~~~ =\frac {1}{\sqrt 2}(|-\rangle_A|+\rangle_B-|+\rangle_A|-\rangle_B),
  \end{eqnarray*}
where
\begin{eqnarray*}
&&|+\rangle=\frac {1}{\sqrt 2}(|0\rangle+|1\rangle),\\
&&|-\rangle=\frac {1}{\sqrt 2}(|0\rangle-|1\rangle).
\end{eqnarray*}
 Obtaining these states could have come about in many different ways; for example, Alice
could prepare
 the pairs and then send half of each to Bob, or vice versa. Alternatively, a third party could prepare the pairs
 and send halves to Alice and Bob. Or they could have met a long time ago and shared them,
 storing them until the present.  Alice and Bob then select a random subset of EPR pairs, and test to see
 if they violate Bell's inequality, or some other appropriate test of fidelity.  Passing the test certifies
 that they continue to hold sufficiently pure, entangled quantum states.  However,  if tampering has occurred,
 Alice and Bob discard  these EPR pairs, and new EPR pairs should be constructed again.
  Without loss of generality we suppose that all  the EPR pairs used in our scheme
are  the Bell state $|\Phi^+\rangle_{AB}$.

{\it Secure direct communication using teleportation} --- After insuring the security of the quantum channel
(EPR pairs), we begin secure direct communication.  Suppose that Bob has a particle sequence and he wishes to
communicate information to Alice. First Bob makes his  particle sequence in the states, composed of $|+\rangle$
and $|-\rangle$, according to the message sequence. For example if the message  to be transmitted is 01001, then
the sequence of particle states  should be in the state $|+\rangle|-\rangle|+\rangle|+\rangle|-\rangle$, i.e.
$|+\rangle$ and $|-\rangle$ correspond to 0 and 1 respectively. Remarkably quantum entanglement of EPR pairs can
serve as a channel for transmission of messages encoded in the sequence of particle states. This is the process
so called quantum teleportation \cite {s23} which we now describe.  We will use subscripts $A$ and $B$ for the
systems which comprise $|\Phi^+\rangle_{AB}$ and the subscript $C$ for Bob's particles with messages. The
systems $B$ and $C$ are thus in Bob's possession and $A$ is in Alice's possession. In components we write the
qubit state carrying message
\begin{equation*}
|\Psi\rangle_C=\frac {1}{\sqrt 2}(|0\rangle_C+b|1\rangle_C),
\end{equation*}
where $b=1$ and $b=-1$ correspond to $|+\rangle$ and $|-\rangle$ respectively.   The overall state of the three
qubits is
\begin{eqnarray*}
&&~~~|\Phi^+\rangle_{AB}|\Psi\rangle_C\\
&&=\frac
{1}{ 2}(|00\rangle_{AB}+|11\rangle_{AB})(|0\rangle_C+b|1\rangle_C)\\
&&=\frac {1}{2\sqrt 2}\{(|0\rangle_A+b|1\rangle_A)|\Phi^+\rangle_{BC}+(|0\rangle_A-b|1\rangle_A)|\Phi^-\rangle_{BC}\\
&&~~~~ +(b|0\rangle_A+|1\rangle_A)|\Psi^+\rangle_{BC} +(b|0\rangle_A-|1\rangle_A)|\Psi^-\rangle_{BC})\}.
\end{eqnarray*}
Thus if Bob performs a Bell measurement on his two particles $BC$ then regardless of the identity of
$|\Psi\rangle_C$, each outcome will occur with equal probability $\frac {1}{4}$. Hence after this measurement,
the resulting state of Alice's particle  will be respectively
\begin{eqnarray*}
&&\frac {1}{\sqrt 2}(|0\rangle_A+b|1\rangle_A)=U_{00}|\Psi\rangle_A,\\
&&\frac {1}{\sqrt 2}(|0\rangle_A-b|1\rangle_A)=U_{01}|\Psi\rangle_A,\\
&&\frac {1}{\sqrt 2}(b|0\rangle_A+|1\rangle_A)=U_{10}|\Psi\rangle_A,\\
&&\frac {1}{\sqrt 2}(b|0\rangle_A-|1\rangle_A)=U_{11}|\Psi\rangle_A,
\end{eqnarray*}
where $U_{ij}$ are
\begin{equation*}
\begin{array}{ll}
U_{00}=(\begin{array}{cc} 1&0\\
0&1\end{array}), &  U_{01}=(\begin{array}{cc}1&0\\
0&-1\end{array}),\\
 U_{10}=(\begin{array}{cc} 0&1\\
1&0\end{array}), & U_{11}=(\begin{array}{cc} 0&1\\
-1&0\end{array}). \end{array}
\end{equation*}
 The two bits $ij$ label the four possible outcomes of the Bell
measurement. Now the crucial observation is that in each case the state of Alice's particle is related to
$|\Psi\rangle_C$ by a fixed unitary transformation $U_{ij}$ independent of the identity of $|\Psi\rangle$. Thus
if Bob communicates to Alice the two bits $ij$ of classical information  (i.e. his actual Bell measurement
outcome) in a public channel then Alice will be able to apply the corresponding inverse transformation
$U_{ij}^{-1}$ to her particle, restoring it to state $|\Psi\rangle_A$ in every case. Then Alice measures  the
base $\{|+\rangle, |-\rangle\}$ and  read out the  messages that Bob want to transmit to her.

  This process of quantum teleportation has various notable features. Once Alice and Bob are in the
possession of their shared entanglement EPR pair, it is entirely unaffected by any noise in the spatial
environment between them. Thus teleporattion achieves perfect transmission of delicate  information across a
noisy environment assuming that classical information is robust and easy to protect against noise (as it is).
Also the entanglement of EPR pair is independent of the spatial location of Alice relative to Bob so that Alice
may travel around and Bob  can transfer the  information without even knowing her location -- he needs only to
broadcast the two bit information of his Bell measurement outcome say, by publishing it in a newspaper
advertisement. In the process of teleportation, Alice is left with a perfect instance of $|\Psi\rangle$ and
hence no participants can gains any further information about its identity. So in our scheme  teleportation
transmits Bob's message without revealing any information to a potential eavesdropper if the quantum channel is
 perfect EPR pairs (perfect quantum channel).

The security  of this protocol only depends on the perfect  quantum channel (pure EPR pairs). Thus as long as
the quantum channel is perfect, our scheme is secure and confidential. By means of the schemes testing the
security of quantum channel in Refs. \cite {s2, s13, s22}, we can ensure that the quantum channel is perfect. So
our  scheme for direct communication using EPR pairs and teleportation is absolutely reliable, deterministic and
secure.

We should pointed out that it is necessary for testing the security of quantum channel, since a potential
eavesdropper may obtain information as following:

(1) Eve can use the entanglement pair to obtain information. Suppose that Eve has a particle pair in the state
$|\Phi^+\rangle_{DE}$. When Eve obtains particle $B$ in preparing  EPR pair, she performs a Bell measurement on
the particles $BD$. Then the particles $AE$ will be in one of the entanglement states $\{|\Phi^+\rangle_{AE},
|\Phi^-\rangle_{AE}, |\Psi^+\rangle_{AE}, |\Psi^-\rangle_{AE} \}$. The entanglement states will be determined by
the measurement outcome  according to the following equation
\begin{eqnarray*}
&&~~~|\Phi^+\rangle_{AB}|\Phi^+\rangle_{DE}\\
&& =\frac {1}{2}(|\Phi^+\rangle_{BD}|\Phi^+\rangle_{AE}+|\Phi^-\rangle_{BD}|\Phi^-\rangle_{AE}
\\&&~~+|\Psi^+\rangle_{BD}|\Psi^+\rangle_{AE}+|\Psi^-\rangle_{BD}\rangle|\Psi^-\rangle_{AE}).
\end{eqnarray*}
Suppose after the measurement the state of particles $BD$ collapses to the state $|\Phi^-\rangle_{BD}$, thus the
particles $AE$ must be in the state $|\Phi^-\rangle_{AE}$. Then Eve will transmit the particle $B$ to Bob. Both
Alice and Bob does not know that there is a potential eavesdropper listening to their conversation  if they do
not test the quantum channel. Bob will proceed  as usual. Therefore a part of  messages might be leaked to Eve.

However by testing quantum channel we can find Eve and avoid the information being leaked. In fact after the
Bell measurement performed by Eve, particles $AE$ are in an entangled state
 \begin{eqnarray*}
&& |\Phi^-\rangle_{AE}=\frac {1}{\sqrt 2}(|+\rangle_A|-\rangle_E+|-\rangle_A|+\rangle_E),\\
\end{eqnarray*}
 and particles $BD$ are also in an  entangled state
 \begin{eqnarray*}
&& |\Phi^-\rangle_{BD} =\frac {1}{\sqrt 2}(|+\rangle_B|-\rangle_D+|-\rangle_B|+\rangle_D),
\end{eqnarray*}
but there is not any correlation between $A$ and $B$.
  So when Alice and Bob perform the measurement
 in the base $\{|+\rangle, |-\rangle\}$ independently, the result will be random without  any correlation. If it is the case
 we can assert that an  eavesdropper exist  and the EPR pairs should be discarded.

 (2) Eve can obtain information by  coupled EPR pair with her probe in  preparing EPR pair.
 We can test whether the quantum channel is perfect or not in this case by the following strategy.
  We select a random  subset of EPR pairs. Alice and Bob perform a measurement in  base $\{|0\rangle, |1\rangle\}$ or
   base $\{|+\rangle, |-\rangle\}$ randomly. If the measurement outcomes  are completely correlation in the same base of Alice and Bob,
   then the quantum channel is completely  perfect or secure, because EPR pair state is the simultaneous
   eigenstate of the operators $\sigma_x^A\sigma_x^B$ and $\sigma_z^A\sigma_z^B$ with the same eigenvalue 1.
   Here $\sigma_x$ and $\sigma_z$ are Pauli operators.
   However if the measurement outcomes of Alice and Bob are  not correlation completely in the same base chosen by Alice and Bob,
   there might be a potential Eve, who have coupled EPR pair with her probe.
   Here we omit the proof and give an example of this case only. Let  Alice's particle $A$, Bob's particle $B$ and Eve's particle
 $F$  in the following entangled state
 \begin{eqnarray*}
&&~~~|\Phi^+\rangle_{ABF}=\frac {1}{\sqrt 2}(|000\rangle+|111\rangle)\\
&&=\frac {1}{2}[|+\rangle_A(|+\rangle_B|+\rangle_F+|-\rangle_B|-\rangle_F)\\
&&+|-\rangle_A(|+\rangle_B|-\rangle_F+|-\rangle_B|+\rangle_F)].
 \end{eqnarray*}
If Alice and Bob perform a measurement in the base $\{|+\rangle, |-\rangle\} $ whether Alice's measurement
outcome is  $|+\rangle_A$ or  $|-\rangle_A$, Bob will have  the same probability $\frac {1}{2}$ to obtain
$|+\rangle_B$ and $|-\rangle_B$. That is to say that Alice's outcome is not correlation with that of Bob's. If
this is the case, evidently there is a potential eavesdropper. We should abandon the quantum channel.

However, under any case, as long as eavesdropper exists, we can find her and insure the security of quantum
channel to realize secure direct communication.

In summary, we give a scheme for secure  direct communication. There is no need for establishing a shared secret
key in this protocol. The communication is based on Einstein-Podolsky-Rosen pairs and teleportation between
Alice and Bob. After insuring the security of the quantum channel (EPR pairs), Bob encodes the secret message
directly on a sequence of  particle states  and transmits them  to Alice by teleportation. Evidently
teleportation transmits Bob's message without revealing any information to a potential eavesdropper. Alice can
read out the encoded messages directly by the measurement on her qubits. Because there is not a transmission of
the qubit which carries  the secret message between Alice and Bob, it is completely secure for direct secret
communication if perfect quantum channel is used.

 In the scheme of Deng et al \cite {s22},  the security  of quantum channel  and  without
 revealing any information to a potential
eavesdropper in transferring information has been guaranteed. However  if Eve intercepts the particles carrying
secret messages in the transmitting channel, although she can not obtain any information, she can make
interruption of communication. In our scheme  information was transmitted using  teleportation, the
communication  can not be  intercepted. Therefore our new protocol has high capacity to defect interference.

In Ekert's QKD protocol \cite {s2} only half of EPR pairs are contributed  to  a secret key. In our scheme all
perfect EPR pairs may be used to transmit  information. Obviously in our protocol the entanglement resources,
the EPR pairs, are used much more efficiently.

Teleportation has been realized in the experiments \cite {s24, s25, s26}, therefore our protocol for secure
direct communication will be realized by  experiment easily.

\begin{acknowledgements}
 This work was supported  by  Hebei Natural Science
Foundation under Grant No. 101094.
\end{acknowledgements}

\end{document}